\documentclass[letterpaper,twocolumn]{IEEEtran}
\usepackage{color}
\usepackage{amssymb}
\usepackage{amsmath}
\usepackage{fixltx2e} 		% Package for subscripts in text mode using \textsubscript
\usepackage{graphicx}
\usepackage{lineno} 			%\linenumbers
\usepackage{siunitx}
\usepackage{amsthm}
\usepackage{epstopdf}
\usepackage{tocloft}
\usepackage{todonotes}
\author{\IEEEauthorblockN{Shi Zhao and David A. Howey \\}
\IEEEauthorblockA{
Department~of~Engineering~Science\\
University~of~Oxford\\
Oxford, United~Kingdom}\\
shi.zhao@eng.ox.ac.uk; david.howey@eng.ox.ac.uk
\thanks{This work is funded through the RCUK Energy Programme's STABLE-NET project (ref. EP/L014343/1).}}

\title{Global sensitivity analysis of battery equivalent circuit model parameters}
\date{Monday, March 14, 2016}

\begin{document}
\maketitle
\begin{abstract}
This paper considers one of the most commonly used equivalent circuit models (ECMs) for lithium-ion batteries and investigates the sensitivity of the model output to changes of model parameters using the Morris method. Experiments are carried out on a lithium-ion cell with nickel manganese cobalt oxide (NMC) electrode and parameters of the model are identified in the state of charge (SOC) range $[100\%,10\%]$. Although all the model parameters do vary with SOC, global sensitivity analysis reveals that the uncertainties of some of the parameters generate very little uncertainty in the voltage output, implying that those parameters can be taken as constants without compromising the accuracy of the model. This is further confirmed by experimental validation.
\end{abstract}


\section{Introduction}
Equivalent circuit models (ECMs) for lithium-ion batteries are widely used for state of charge (SOC) and state of health (SOH) estimation in battery management systems (BMSs) \cite{hu2012comparative}. Compared to electrochemical models which are derived from electrochemical principles and are characterised by partial differential equations coupled with algebraic equations (PDAEs), ECMs are low order models parameterised from time-domain or frequency domain experimental data and are computationally more efficient. In the system identification step, the parameters of the ECMs are adjusted so that the model output matches the experimental measurements as closely as possible. % and the physical meanings of the parameters are somewhat ambiguous.
It is recognised in the literature that the model parameters vary with SOC (and temperature\footnote{In this paper we will focus on the isothermal scenario, thus will not discuss the temperature dependence.}). Therefore in order to improve the model accuracy, they are often taken as functions of SOC \cite{hu2012robustness}. Alternatively, the parameters are set to be constants so that the complexity of the model is reduced \cite{zhao2015observability}.

The choice of whether the ECM parameters should be SOC dependent is a tradeoff between model accuracy and efficiency. And this seems to be a binary choice from the literature: either we vary all the parameters with SOC or we set all the parameters to constants. Such a dilemma, which is implicitly based on the assumption that all the parameters are equally important to the model output, is probably unnecessary as the output may be more sensitive to changes of some parameters than the others. This problem can be resolved by a parameter sensitivity analysis, in which the relative importance of the parameters can be quantified.

Sensitivity analysis also has valuable implications for parameter estimation as it reveals which parameters we have to make the most efforts to estimate. On the other hand, the less important parameters may only need to be estimated roughly since their uncertainties have little impact on the output. This is particularly important for complex models with many parameters.

In this paper, we apply the Morris method, which is a technique for global sensitivity analysis, to the second order RC model for batteries. Experiments are carried out on a lithium nickel manganese cobalt oxide (NMC) battery cell and the experimental data is used to identify the model parameters. It can be seen that some of the parameters change significantly when the cell is discharged from $100\%$ to $10\%$ SOC. However, large change of a parameter in terms of magnitude does not necessarily mean that it has to be taken as a function of the SOC, as demonstrated by the sensitivity analysis.
 %Specifically, the cell is discharged by $10\%$ SOC using a driving cycle and then rested for hours

\section{Morris method}
Consider a general dynamical system described by the state space model
\begin{subequations}
\begin{eqnarray}
  \dot{x} &=& f(x,u;\theta) \\
  y &=& g(x,u;\theta)
\end{eqnarray}
\end{subequations}
where $x \in \mathbb{R}^n$ is the state vector, $u \in \mathbb{R}^p$ is the input, $y \in \mathbb{R}$ is the output\footnote{There is no loss of generality by considering a single-output system since the analysis in the paper can be readily applied to multiple-output systems.} and $\theta \in \mathbb{R}^{q}$ is the parameter set. The objective of parameter sensitivity analysis is to determine how the output $y$ is affected by changes in the parameters $\theta$. Suppose the initial condition $x_0$ and the input $u(t)$ are both fixed, with a slight abuse of notation, we will omit $u$ in the following analysis when there is no risk of confusion. Therefore the output $y$ at time $t$ can be written as $y(t;\theta)$ and the aim is to analyse the sensitivity of $y(t;\theta)$ with respect to $\theta$.

A common mistake for sensitivity analysis is to simply calculate the partial derivative of $y(t;\theta)$ with respect to the $i_{th}$ parameter $\theta_i$, \emph{i.e.}, $\frac{\partial y}{\partial \theta_i}$ around a particular point. This approach is problematic because it does not take the variance of the parameter $\theta_i$ into account. As a simple example, consider the linear function
\begin{equation}%\label{}
y = \theta_1+5\theta_2
\end{equation}
where $\theta_1 \sim \mathcal{N}(0,10)$, $\theta_2 \sim \mathcal{N}(0,1)$ and $\mathrm{Cov}(\theta_1,\theta_2)=0$. Although $\frac{\partial y}{\partial \theta_1} < \frac{\partial y}{\partial \theta_2}$, $y$ is in fact more sensitive to change in $\theta_1$. In other words, the uncertainty of $\theta_1$ contributes more to the uncertainty of $y$.

Therefore a more appropriate way is to compute $\sigma_{\theta_i}\cdot\frac{\partial y}{\partial \theta_i}$ where $\sigma_{\theta_i}$ is the standard deviation of $\theta_i$. This is known as ``one-factor-at-a-time'' (OAT) method \cite{saltelli2006sensitivity}. Although this method can deal with strictly linear systems well, it is still problematic for nonlinear systems since effectively the nonlinear function $y(t;\theta)$ is linearised. Therefore the OAT method is local in nature. Many systems of interest, including the battery ECM we will study in the next section, are highly nonlinear with respect to the parameters.

%Compared to the OAT method, which is local in nature,

Based on the OAT approach, the Morris method achieves its global nature by repeating the local sensitivity analysis multiple times with randomly sampled linearisation points \cite{morris1991factorial}. In the first step of each run, a starting point $\boldsymbol{\theta}$ is obtained by sampling each of the components $\boldsymbol{\theta}_i$ from the normal distribution $\theta_i \sim \mathcal{N}(\mu_i,\sigma_i)$ where $\mu_i$ and $\sigma_i$ are the mean value and the  standard deviation of the $i_{th}$ parameter, respectively\footnote{A more rigorous way is to sample the vector $\boldsymbol{\theta}$ directly if the covariance matrix of the parameters is known.}. In the following steps, the sensitivity of the output to the change of the $i_{th}$ parameter can then be calculated by
\begin{equation}%\label{}
\xi_i = \frac{y(t;\boldsymbol{\theta})-y(t;\boldsymbol{\hat{\theta}_i})}{\Delta}
\end{equation}
where $\Delta$ is a constant, the varied parameter $\boldsymbol{\hat{\theta}_i}$ is obtained by changing the $i_{th}$ entry of $\boldsymbol{\theta}$ to $\boldsymbol{\theta}_i+\Delta \sigma_i$ and keeping all the other parameters unchanged, that is, $\boldsymbol{\hat{\theta}_i} = [\boldsymbol{\theta}_1, \cdots,   \boldsymbol{\theta}_{i-1}, \boldsymbol{\theta}_i+\Delta \sigma_i, \boldsymbol{\theta}_{i+1},  \cdots, \boldsymbol{\theta}_q]$. It is clear so far such a single run is still a local analysis. In the Morris method, the process is repeated $N$ times and the effect of $\theta_i$ can then be seen from the distribution of $\xi_i$ in the $N$ runs. We are usually interested in its mean value. Sometimes, in order to avoid that $\xi_i$ in different runs cancels each other out due to opposite signs, its absolute value is taken instead and this is called the enhanced Morris method \cite{campolongo2005enhancing}.

In essence, the Morris method is a Monte Carlo method that aims to find the probability distribution of the sensitivity of model output to changes of the parameters. The local analysis using the OAT approach in each run gives only a sample of the probability distribution.

We note that there are other techniques for global sensitivity analysis, but it is beyond the scope of this paper to give a comprehensive review. We refer interested reader to \cite{saltelli2006sensitivity} for more details. It should also noted that the analytical form of $y(t;\theta)$ is usually not available and has to be solved numerically.

%Therefore the result of parameter sensitivity analysis using the Morris method

%In the second step, we obtain a varied parameter vector by changing the first entry of $\boldsymbol{\theta}_i$ to $\boldsymbol{\theta}_1+\Delta \sigma_1$ where $\Delta$ is a constant and keeping all the other parameters unchanged, that is, $\boldsymbol{\hat{\theta}_1} = [\boldsymbol{\theta}_1+\Delta \sigma_1, \boldsymbol{\theta}_2 \cdots, \boldsymbol{\theta}_q]$. The model output corresponding to $\boldsymbol{\hat{\theta}_1}$ is calculated. The sensitivity of the output to the change of the first parameter can then be calculated by
%\begin{equation*}%\label{}
%\xi_1 = \frac{y(t;\boldsymbol{\theta})-y(t;\boldsymbol{\hat{\theta}_1})}{\Delta}
%\end{equation*}
%Similarly,

\section{Battery ECM parameterisation}

\subsection{Battery ECM}
\begin{figure}
\centering
\includegraphics[width=0.35\textwidth]{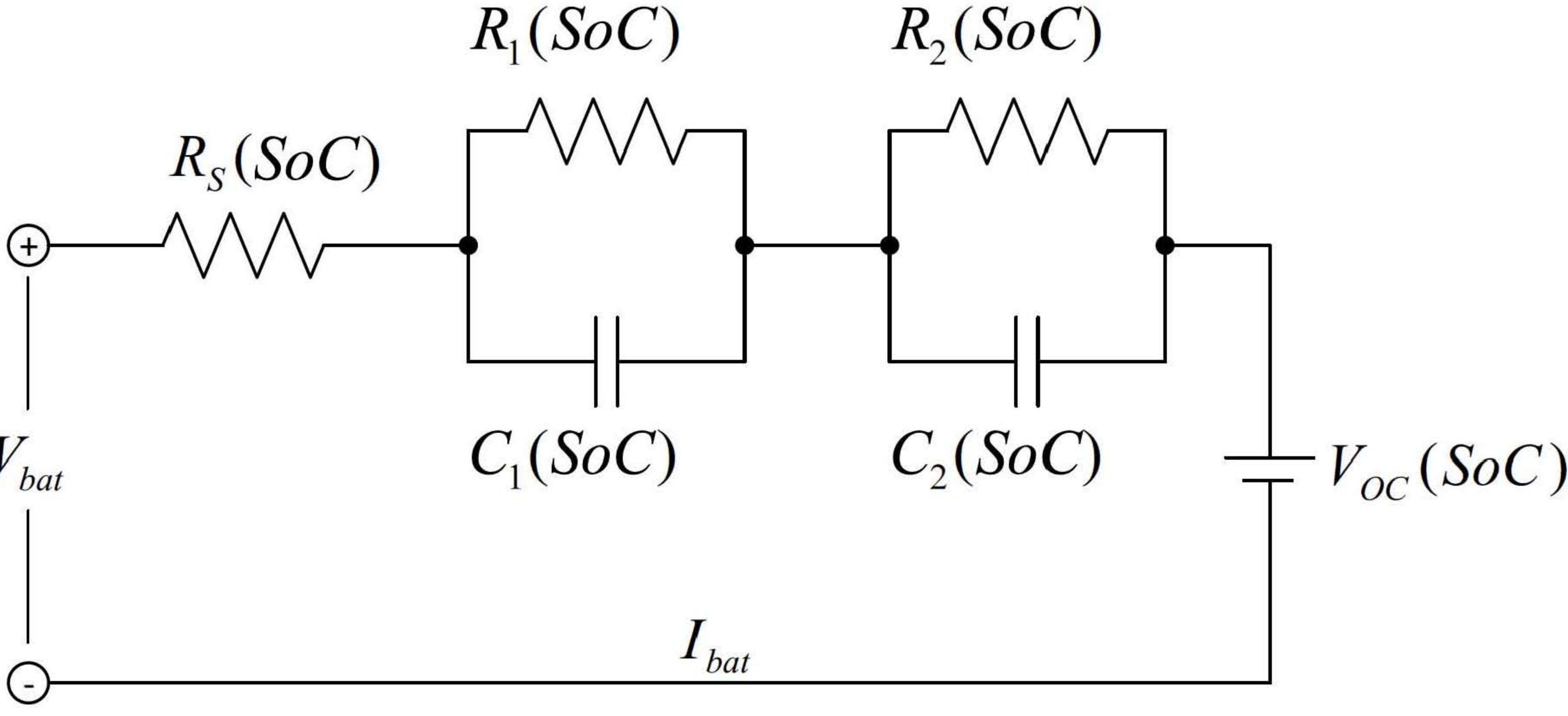}
\caption{Schematic diagram of the ECM.}
\label{fig:Figure1}
\end{figure}

The ECM under consideration here is the second-order RC model shown in Fig. \ref{fig:Figure1}. This standard model has been studied extensively for the purposes of state and/or parameter estimation in the literature. However, to the best of our knowledge, global sensitivity analysis of the parameters in this model, and in other battery models in general, has not been studied before. It is therefore our intention to carry out such an analysis in this paper.

Apart from the open circuit voltage (OCV) $V_{OC}$, which is a nonlinear function of the SOC, there are five parameters in the model: $R_1,R_2,C_1,C_2$ and $R_s$.
 The state space representation of the model in the continuous form is given by
\begin{subequations}\label{sys:02}
\begin{eqnarray}
  \begin{bmatrix} \dot{V}_1 \\ \dot{V}_2 \\ \dot{Z} \end{bmatrix} &=&  \begin{bmatrix} -\frac{1}{R_1 C_1} & 0 & 0 \\ 0 & -\frac{1}{R_2 C_2} & 0 \\ 0 & 0 & 0 \end{bmatrix} \begin{bmatrix} V_1 \\ V_2 \\ Z \end{bmatrix} + \begin{bmatrix} \frac{1}{C_1} \\ \frac{1}{C_2} \\ -\frac{1}{Q}\end{bmatrix} I  \label{sys:02a} \nonumber \\
  \\
  V &=& V_{OC}(Z) - V_1 - V_2 - I R_s \label{sys:02b}
\end{eqnarray}
\end{subequations}
where $V_1,V_2$ are the voltages across the first and the second RC pairs, respectively, $Z \in [\SI{0}{\percent},\SI{100}{\percent}]$ is the normalised SOC, $Q$ is the battery capacity, $I$ is the current, whose sign is taken to be positive when the battery is discharging and negative during charging, and $V$ is the terminal voltage.

It is common to define $\tau_1 = R_1 C_1$ and $\tau_2 = R_2 C_2$ where $\tau_1$ and $\tau_2$ are the time constants of the two RC pairs, respectively. In the parameter estimation step, one can choose to identify either the set $\{R_1,R_2,C_1,C_2, R_s\}$ or the set $\{\tau_1,\tau_2,C_1,C_2, R_s\}$. In this paper, we choose the latter set since it is easier to interpret the physical meanings of time constants.

\subsection{Battery test}
To conduct parameter sensitivity analysis of the model, it is necessary to first determine the ranges of variations of the parameters. This is achieved by experimental investigations and a Kokam NMC lithium-ion pouch cell (SLPB533459) with a nominal capacity 740 mAh was studied. A BioLogic SP-150 potentiostat was used to collect the data with the temperature being fixed at $\SI{20}{\celsius}$ using a thermal chamber.

\begin{figure}
\centering
\includegraphics[width=0.44\textwidth]{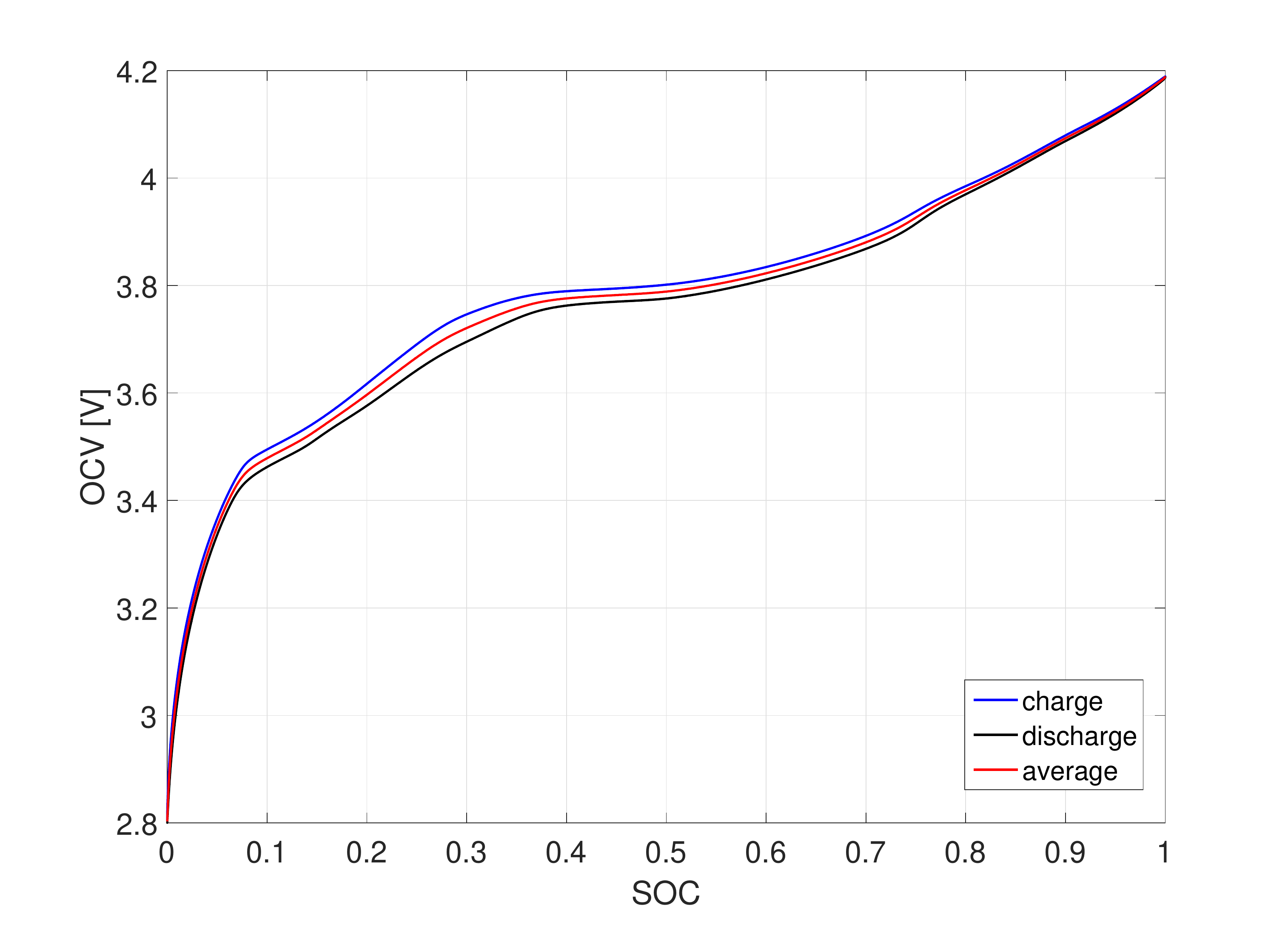}
\caption{OCV as a function of SOC.}
\label{fig:Figure2}
\end{figure}

The relationship between the OCV and SOC was determined by charging and discharging the cell at $1/25$C ($\SI{29.6} {\milli\ampere}$).
%a galvanostatic intermittent titration technique (GITT) procedure in which 50 data points were recorded in both the charge and discharge processes with current 0.1C \cite{birkl2015parametric}.
Due to hysteresis, there is a discrepancy between the charge and discharge curves, as shown in Fig. \ref{fig:Figure2}. The average of these two curves is taken as the OCV-SOC function and is approximated by a polynomial $V_{OC} = \sum_{k=0}^{10} a_k Z^{k}$
%\begin{equation}
%  V_{OC} = \sum_{k=0}^{10} a_k Z^{k}
%\end{equation}
where the coefficients are given in Table \ref{tab:01}. The maximal approximation error of this polynomial in the SOC range $[10\%,100\%]$ compared to the averaged OCV-SOC curve is $\SI{3.92} {\milli\volt}$ and the root-mean-square error (RMSE) is $\SI{1.47} {\milli\volt}$.

%\begin{table}
%\centering
%\begin{tabular}{|c|c|c|c|c|c|c|c|c|c|c|c|c|c|}
%\hline
% coefficient & $a_0$ &$a_1$ & $a_2$ & $a_3$ & $a_4$ & $a_5$ & $a_6$ & $a_7$ & $a_8$ & $a_9$ & $a_{10}$ & $a_{11}$ & $a_{12}$  \\
%  \hline
% value & $2.826756095452177$ &$24.086870497958735$ & $-4.190174510316422e+02$ & $4.283854470347601e+03$ & $-2.731296402930840e+04$ & $1.158755780603406e+05$ & $-3.383907046931966e+05$ & $6.881649454618373e+05$ & $-9.698941354902880e+05$ & $9.265895825848157e+05$ & $-5.714368352006432e+05$ & $2.049611301357813e+05$ & $-3.244416198168942e+04$  \\
%  \hline
%\end{tabular}
%\caption{The values of the OCV polynomial coefficients.}
%\label{tab:02}
%\end{table}

\begin{table}
\centering
\caption{The values of the OCV polynomial coefficients.}
\begin{tabular}{|c|c|}
\hline
 coefficient & value \\
  \hline
  $a_0$ & $2.82$ \\
  $a_1$ & $1.95 \times 10^1$ \\
  $a_2$ & $-2.49 \times 10^2$ \\
  $a_3$ & $1.78 \times 10^3$ \\
  $a_4$ & $-7.47 \times 10^3$\\
  $a_5$ & $1.96\times 10^4$ \\
  $a_6$ & $-3.31\times 10^4$  \\
  $a_7$ & $3.63\times 10^4$ \\
  $a_8$ & $-2.50\times 10^4$ \\
  $a_9$ & $9.77 \times 10^3$ \\
  $a_{10}$ & $-1.66\times 10^3$ \\
  \hline
\end{tabular}
\label{tab:01}
\end{table}

After the OCV-SOC test, the battery cell was fully charged to $100\%$ SOC using a standard constant current constant voltage (CCCV) procedure. Specifically, the cell was charged using a $1$C current ($\SI{740} {\milli\ampere}$) until the voltage reached $\SI{4.2} {\volt}$ and was then charged with the voltage kept at $\SI{4.2} {\volt}$ until the current dropped to $1/25$C ($\SI{29.6} {\milli\ampere}$). The fully charged cell was rested for one hour.

\begin{figure}
\centering
\includegraphics[width=0.52\textwidth]{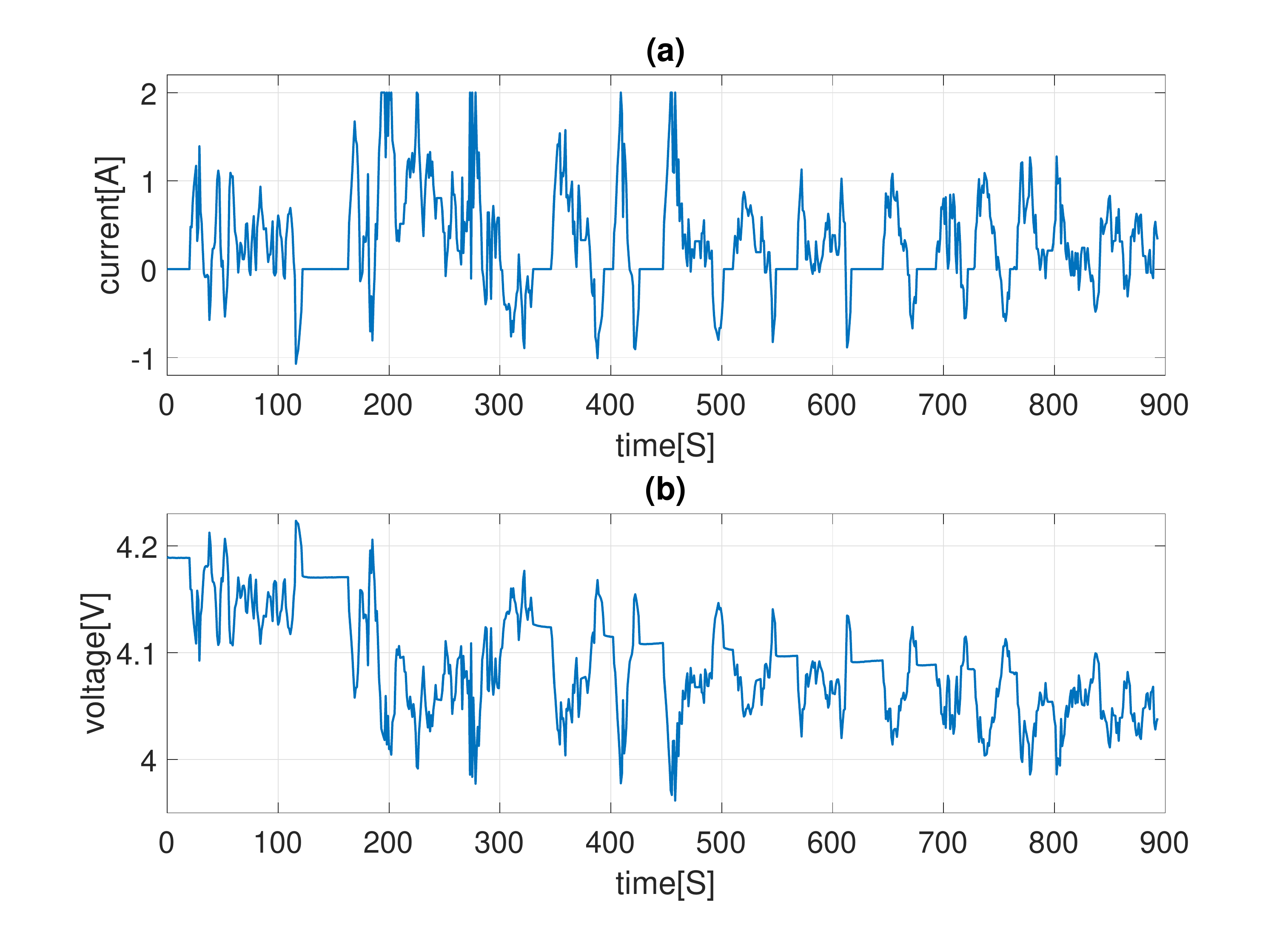}
\caption{(a): The FUDS current profile used to discharge the cell; (b): The terminal voltage of the battery cell.}
\label{fig:Figure3}
\end{figure}

The federal urban driving schedule (FUDS) current profile \cite{testmanual} was then applied to the cell until it was discharged by roughly $10\%$ SOC. Fig. \ref{fig:Figure3} shows the experimental measurements of the input (current) and the output (voltage). After discharging, the cell had about $90\%$ SOC and it was rested for one hour. The same discharging and resting procedure was then repeated until the SOC dropped to about $10\%$.  We note that the SOC was calculated by Coulomb couting using the accurate laboratory equipment.

\subsection{Parameter estimation}
%The model parameters are treated as constants during each of the discharging cycles as the change in SOC is only $10\%$. However, they are expected to vary in different discharging cycles.
The model parameters are identified using MATLAB\textsuperscript {\circledR}'s System Identification Toolbox. The first step is to subtract the OCV, which is calculated using the fitted polynomial $V_{OC}$, from the terminal voltage. This gives a linear state space model where the input is $I$ and the output is $V-V_{OC}$. The initial values of the state vector $[V_1, V_2]^T$ are set to $[0,0]^T$ as the battery has been rested for one hour before the current is applied. Since the model structure is known, we use MATLAB functions \texttt{idgrey} and \texttt{greyest} to estimate the parameter set $\{\tau_1,\tau_2,C_1,C_2, R_s\}$.

\begin{figure}
\centering
\includegraphics[width=0.52\textwidth]{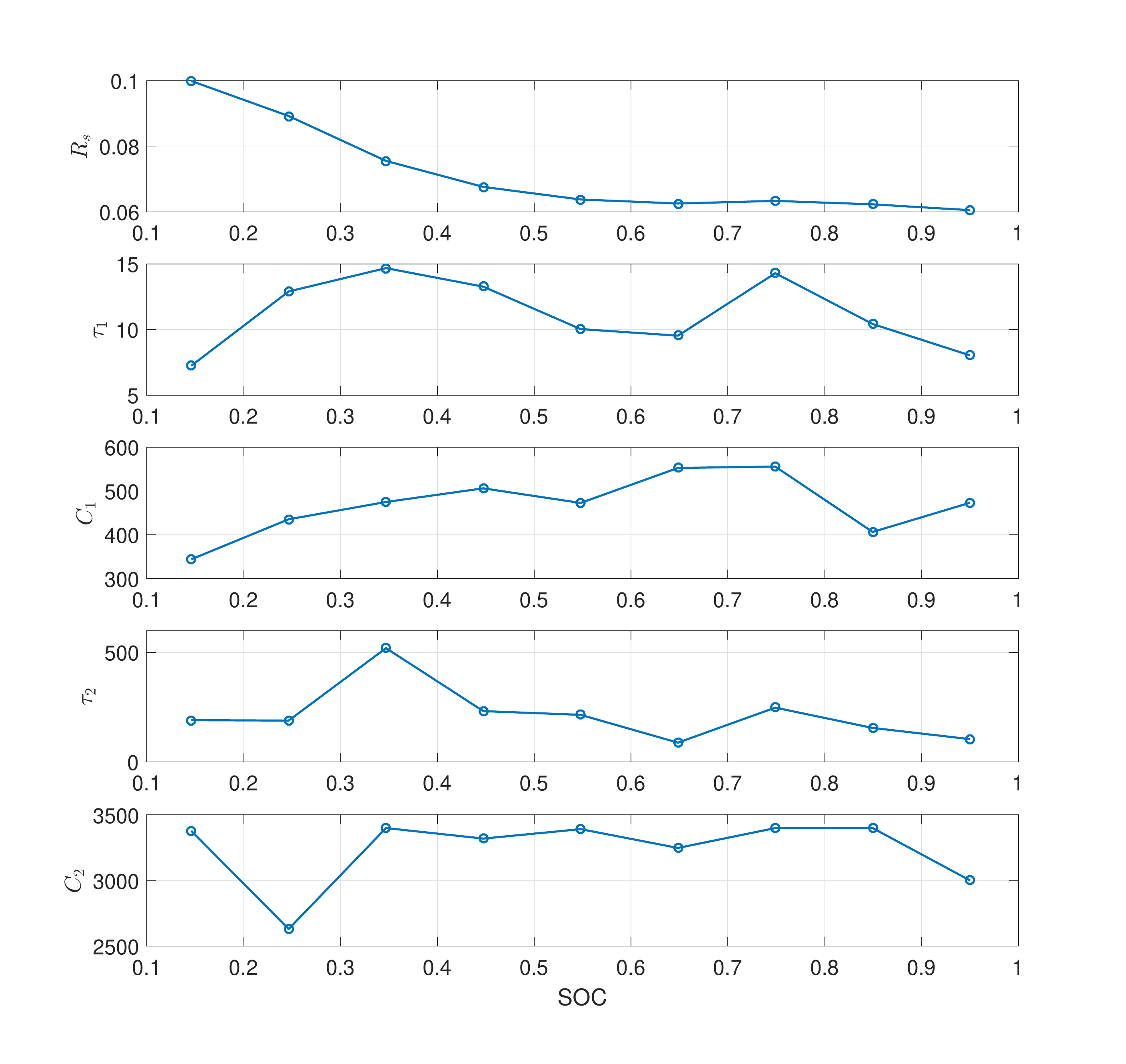}
\caption{ECM parameters obtained by system identification in different SOC intervals.}
\label{fig:Figure4}
\end{figure}

Using the experimental data, we identify the parameters when the SOC is in the intervals $[100\%,90\%), [90\%,80\%),\cdots,[20\%,10\%)$. The parameters are set to be constants in each SOC interval. However, they are allowed to vary in different intervals. Therefore we obtain $9$ sets of parameters, which are shown in Fig. \ref{fig:Figure4}, where the midpoints of the intervals are used in the horizontal axis.
It can be seen from Fig. \ref{fig:Figure4} that all the parameters do vary with SOC. It is common to use polynomials to fit the parameters as functions of SOC. Sometimes, piecewise linear functions are used instead.

%The parameters are treated as constants since the change in SOC is only $10\%$. However, they are expected to vary in different discharging cycles.

\section{Sensitivity analysis}
%Sensitivity analysis addresses the questions that which parameters can be set to be constants in their ranges of variations without compromising the model accuracy much. %It also helps us understand which parameters we should make the most efforts to estimate.

\subsection{Main results}
\begin{figure}
\centering
\includegraphics[width=0.52\textwidth]{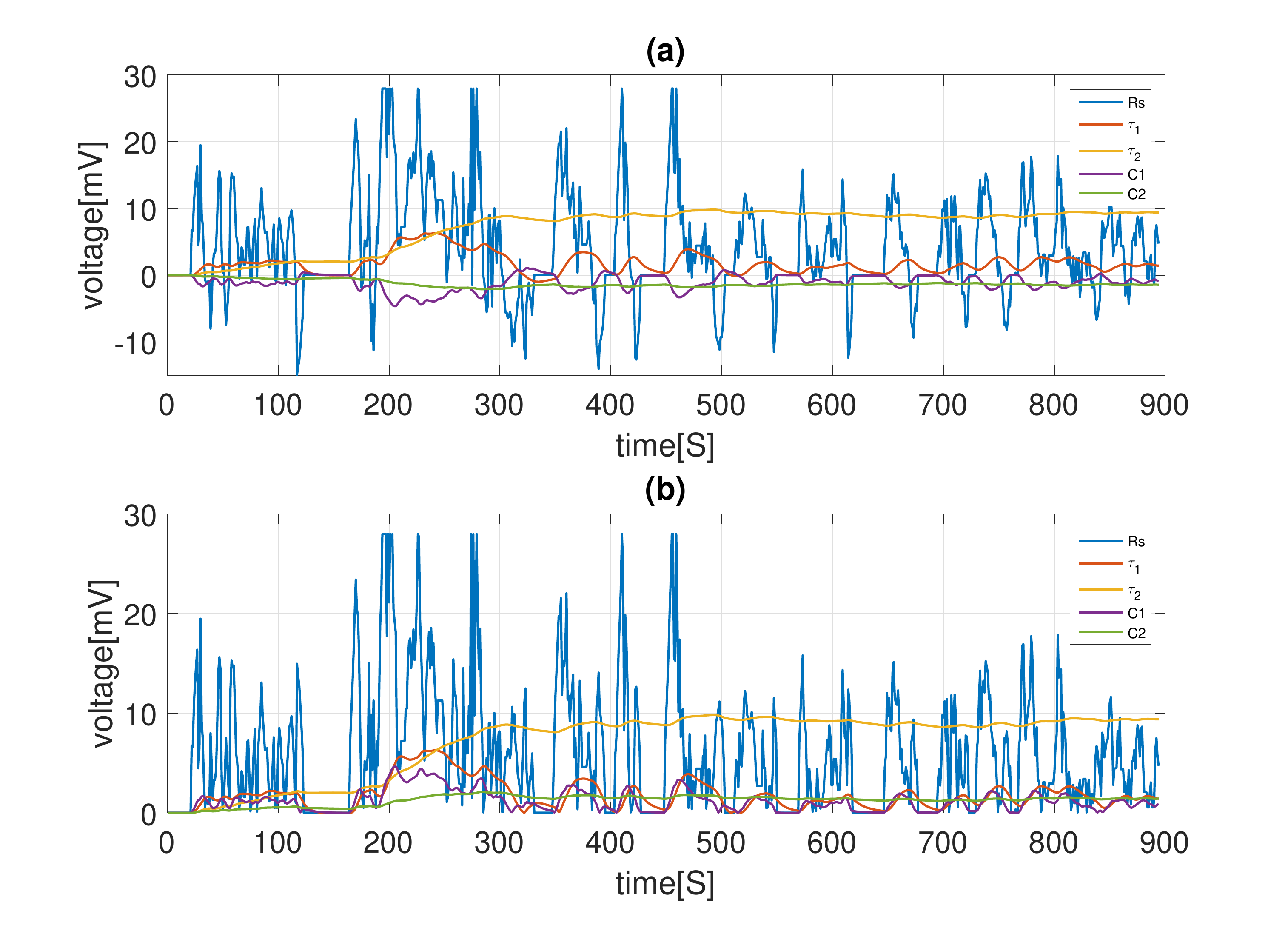}
\caption{Parameter sensitivity analysis results obtained by (a) the Morris method and (b) the enhanced Morris method. }
\label{fig:Figure5}
\end{figure}

The Morris method is used to study the sensitivity of the model output to changes of the parameters. The mean values and standard deviations of the parameters in different SOC intervals are calculated from the results of system identification. All the parameters are considered to be mutually independent. We run $N=1024$ Monte Carlo simulations using randomly sampled starting points and in each simulation, an OAT analysis is carried out. The averages of the results from the $N$ runs are illustrated in Fig. \ref{fig:Figure5}(a), which shows how much the model output changes when a particular parameter is changed by one standard deviation. Fig. \ref{fig:Figure5}(b) shows the results given by the enhanced Morris method. A comparison between Fig. \ref{fig:Figure5}(a) and Fig. \ref{fig:Figure5}(b) implies that the cancellation effects are not severe.

It can be seen from Fig. \ref{fig:Figure5} that the model output is most sensitive to the variation of the series resistance $R_s$. This is not surprising as the output is directly related to $R_s$ in (\ref{sys:02b}). In fact, the sensitivity analysis of the output to $R_s$ can be conducted using the OAT method because $V$ is a linear function of $R_s$ when all the other parameters are fixed. The time constant $\tau_2$ is the second most important parameter in terms of sensitivity. Changing $\tau_2$ by a standard deviation will result in a difference of about $\SI{10} {\milli\volt}$  in the output. The remaining three parameters, $C_1, C_2$ and $\tau_1$ contribute little to the uncertainty of the output, therefore we expect that these parameters can be set to constants if the model complexity needs to be reduced.

\subsection{Experimental validation}
%The conclusion of the analysis still needs to be validated and we are especially interested if it still holds when the input is different. The ARTEMIS European urban driving cycle (UDC) \cite{Andre2004} shown in Fig. \ref{fig:Figure6}(a) was applied to the same fully charged cell until its SOC dropped to less than $20\%$. We compare the experimental voltage measurements to the output of the model in which all the parameters vary with SOC, as shown in Fig. \ref{fig:Figure4}. The maximum  error is $\SI{21.48} {\milli\volt}$ and the RMSE is $\SI{5.67} {\milli\volt}$. If $C_1, C_2$ and $\tau_1$ are fixed at their mean values (case 2 in Fig. \ref{fig:Figure6}(b)), the maximum  error is $\SI{25.91} {\milli\volt}$ and the RMSE is $\SI{7.12} {\milli\volt}$. This confirms that these parameters can be set to constants without affecting the model accuracy much. By contrast, the maximum  error is $\SI{45.52} {\milli\volt}$ and the RMSE is $\SI{15.35} {\milli\volt}$ if all the parameters are fixed at their mean values (case 3 in Fig. \ref{fig:Figure6}(b)).

\begin{figure}
\centering
\includegraphics[width=0.52\textwidth]{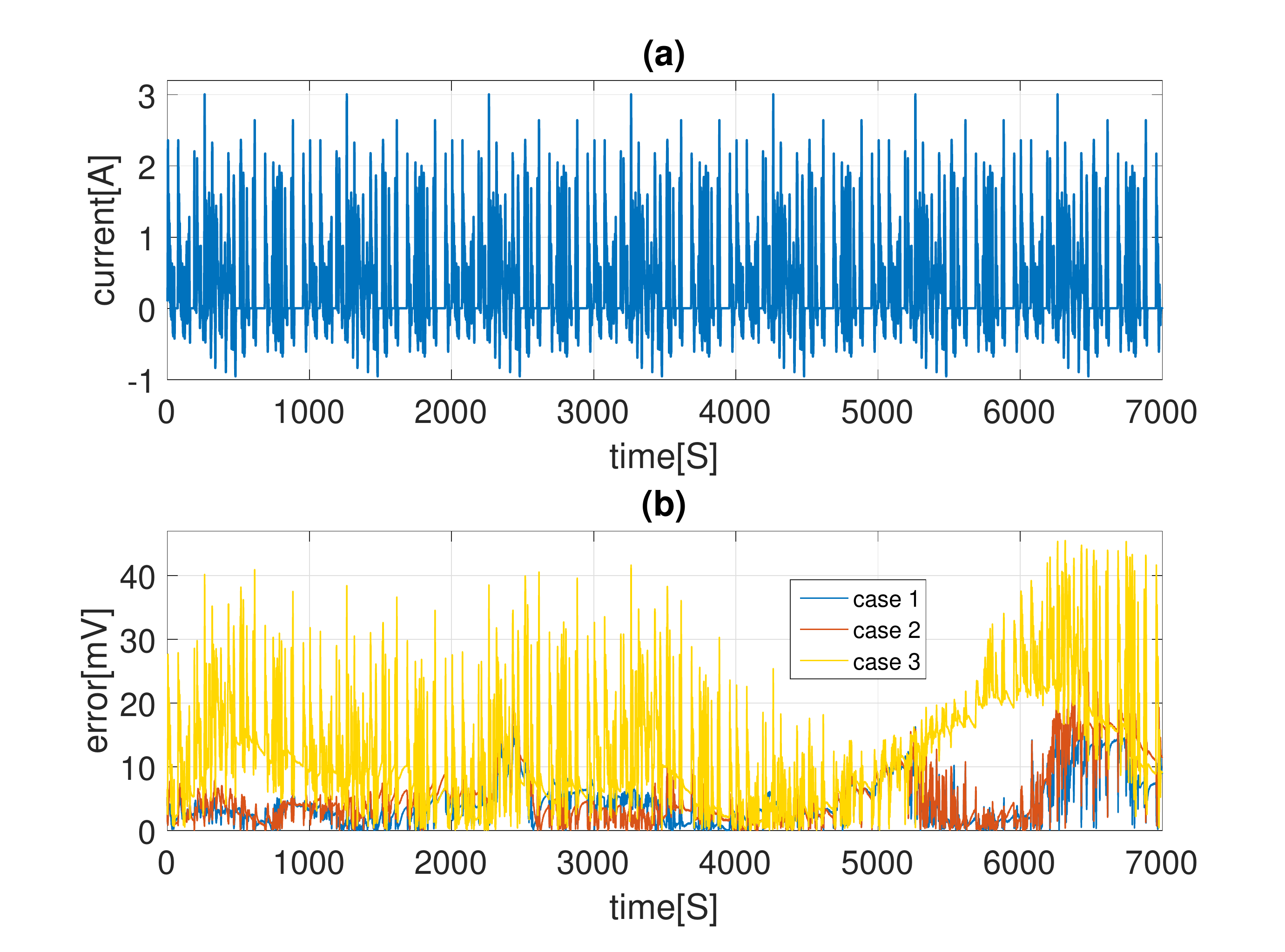}
\caption{Experimental validation. (a): The UDC current profile; (b): Model error in case 1: all the parameters vary with SOC; case 2: $C_1, C_2$ and $\tau_1$ are fixed; case 3: all the parameters are fixed.}
\label{fig:Figure6}
\end{figure}

The conclusion of the analysis still needs to be validated and we are especially interested if it still holds when the input is different. The ARTEMIS European urban driving cycle (UDC) \cite{Andre2004} shown in Fig. \ref{fig:Figure6}(a) was applied to the same fully charged cell until its SOC dropped to less than $20\%$. We compare the experimental voltage measurements to the output of the model in which all the parameters vary with SOC. The error is shown by the blue line in Fig. \ref{fig:Figure6}(b) (case 1). The maximum  error is $\SI{21.48} {\milli\volt}$ and the RMSE is $\SI{5.67} {\milli\volt}$. If $C_1, C_2$ and $\tau_1$ are fixed at their mean values (case 2), the maximum  error is $\SI{25.91} {\milli\volt}$ and the RMSE is $\SI{7.12} {\milli\volt}$. This confirms that these parameters can be set to constants without affecting the model accuracy much. By contrast, the maximum  error is $\SI{45.52} {\milli\volt}$ and the RMSE is $\SI{15.35} {\milli\volt}$ if all the parameters are fixed at their mean values (case 3).

\section{Conclusion}
%In this paper, the Morris method is used to analyse the sensitivity of the output of the second order RC model to the changes of the model parameters.
Sensitivity analysis demonstrates that the output of the second order RC model is most sensitive to $R_s$ and $\tau_2$, while the other three parameters $C_1, C_2$ and $\tau_1$ can be set to constants. Future work includes taking temperature effects into account and extending the analysis into more complex battery models.

%\section*{Acknowledgements}
%The authors would like to thank Christoph Birkl for his help with experiments.

%\bibliographystyle{IEEEtran}
%\bibliography{sensitivity_paper_library}

\end{document}